\documentclass[a4paper]{article}

\usepackage{icrc2013}
\usepackage[english]{babel}

\title{Solar neutrino analysis of Super-Kamiokande}

\shorttitle{Super-Kamionade Solar neutrino}

\authors{
Hiroyuki Sekiya$^{1,2}$
for the Super-Kamiokande Collaboration.
}

\afiliations{
$^1$ Kamioka Observatory, Institute for Cosmic Ray Research, University of Tokyo\\
$^2$ Kavli Institute for the Physics and Mathematics of the Universe (WPI), University of Tokyo\\
\scriptsize{

}
}

\email{sekiya@icrr.u-tokyo.ac.jp}

\abstract{
Super-Kamiokande-IV data taking began in September of 2008, and with upgraded electronics
and improvements to water system dynamics, calibration and analysis techniques, a clear solar
neutrino signal could be extracted at recoil electron kinetic energies as low as 3.5 MeV.
The SK-IV extracted solar neutrino flux between 3.5 and
19.5 MeV is found to be (2.36$\pm$0.02(stat.)$\pm$0.04(syst.))$\times 10^6$ /(cm$^2$sec). 
The SK combined recoil electron energy spectrum favors distortions predicted by standard neutrino flavour oscillation parameters 
over a flat suppression at 1$\sigma$ level.
A maximum likelihood fit to the amplitude of the expected solar zenith angle variation of 
the elastic neutrino-electron scattering rate in SK, results in a day/night asymmetry of 
$-3.2\pm1.1$(stat.)$\pm$0.5(syst.)$\%$. The 2.7 $\sigma$ significance of non-zero asymmetry
is the first indication of the regeneration of electron type solar neutrinos as they travel through Earth's
matter. 
A fit to all solar neutrino data and KamLAND yields
$\sin^2 \theta_{12} = 0.304 \pm 0.013$, $\sin^2 \theta_{13} = 0.031^{+0.017}_{-0.015}$ and 
$\Delta m^2_{21} = 7.45^{+0.20}_{-0.19} \times 10^{-5} {\rm eV}^2$. 
}


\begin{document}
\maketitle

\section{Introduction}
Solar neutrino flux measurements from Super-Kamiokande (SK)~\cite{bib:1} and the Sudbury Neutrino Observatory(SNO)~\cite{bib:2} 
have provided direct evidence for solar neutrino flavor conversion. However, there is still no clear evidence that 
this solar neutrino flavor conversion is indeed due to neutrino oscillations and not caused by any other mechanism.
Currently there are two testable signatures unique to neutrino oscillations. The
first is the observation and precision test of the MSW resonance curve~\cite{bib:4}. 
Based on oscillation parameters extracted from solar neutrino and reactor anti-neutrino
measurements, there is an expected characteristic energy dependence of the flavor conversion. 
The higher energy solar neutrinos (higher energy $^8$B and hep neutrinos) undergo
complete resonant conversion within the sun, while the flavor changes of the lower energy solar neutrinos
(pp, $^7$Be, pep, CNO and lower energy $^8$B neutrinos) arise only from vacuum oscillations, which limits the average
electron flavor survival probability to exceed 50$\%$. The transition from the matter dominated oscillations within
the sun, to the vacuum dominated oscillations, should occur near 3 MeV, making $^8$B neutrinos the best
choice when looking for a transition point within the energy spectrum. 
A second signature unique to oscillations arises from the effect of the terrestrial matter density on
solar neutrino oscillations. This effect is tested directly by comparing solar neutrinos which pass through the
Earth at nighttime to those which do not during the daytime.
Those neutrinos which pass through the Earth will in general have an enhanced electron neutrino content
compared to those which do not, leading to an increase in the nighttime electron elastic scattering rate (or any
charged-current interaction rate), and hence a negative ``day/night asymmetry''.
SK detects $^8$B solar neutrinos over a wide energy range in real time, making it a prime detector
to search for both solar neutrino oscillation signatures.

In this Presentation, the energy spectrum results of SK-IV,
the combined SK day/night asymmetry analysis, and an oscillation analysis of SK data
and a global analysis which combines the SK results with other relevant experiments are presented.

\section{Improvements of Super-Kamiokande IV}
Super-Kamiokande is a large, cylindrical, water
Cherenkov detector consisting of 50,000 tons of ultra pure
water located underground, 1000 m underneath
Mount Ikenoyama, in Kamioka City, Japan. The SK
detector is optically separated into a 32.5 kton cylindrical
inner detector (ID) surrounded by a 2.7 meter active
veto outer detector (OD). The structure dividing
the detector regions contains an array of photo-multiplier
tubes (PMTs). In October of 2006, with 11,129 inner and 1,885
outer PMTs, data taking re-started as the SK-III phase~\cite{bib:6}.
The fourth phase of SK (SK-IV) began in September of 2008, 
with new front-end electronics for both the inner and outer detectors, 
and continues to run. 

Improving the front-end electronics, the water circulation
system, calibration techniques and the analysis
methods have allowed the SK-IV solar neutrino measurement
to be made with a lower energy threshold and with
a lower systematic uncertainty, compared to SK-I, II and III.

The new front-end electronics called QBEEs were installed, allowing for
the development of a new online data acquisition system.
The essential components on the QBEEs, used for the
analog signal processing and digitization, are the QTC
(high-speed Charge-to-Time Converter) ASICs~\cite{bib:7}, which
achieve very high speed signal processing and allow the
recording of every hit of every PMT. The resulting PMT
hits information are sent to online computers where a
software trigger searches for timing coincidences within
200 ns to pick out events. The energy threshold of this software trigger is only limited by the
speed of the online computers. 

Ultra-pure water is continuously supplied from the bottom
of the detector and drained from the top, as it is circulated through the water purification 
system with a flow rate of 60 ton/hour. If a temperature
gradient exists within the detector and the supply water temperature is different from 
the detector temperature, convection can occur
throughout the detector volume and radioactive radon gas, which is usually produced by decays from the U/Th
chain near the edge of the detector, can make its way into the center of the detector. Radioactivity coming from
the decay products of radon gas, most commonly $^{214}$Bi beta decays, 
can mimic the lowest energy solar neutrino events. In January of 2010, a new automated temperature
control system was installed, allowing for control of the supply water temperature at the $\pm0.01$ K level.
By controlling the water flow rate and the supply water
temperature (within 0.01 K), convection within the tank
can be kept to a minimum and the background level in the central region of the fiducial volume has since become
significantly lower.

In addition to hardware improvements, a new analysis method had been developed.
Even at the low energies of the $^8$B solar neutrinos, it is possible to use the PMT hit pattern of the Cherenkov
cones to reconstruct the multiple Coulomb scattering of the resultant electrons. Very low energy electrons will
incur more multiple scattering than higher energy electrons and thus have a more isotropic PMT hit pattern.
The radioactive background events such as $^{214}$Bi beta decays generally have less energy than $^8$B solar neutrinos.
To characterize this hit pattern anisotropy, a ``direction fit '' goodness is used. This goodness is constructed
by first projecting 42$^{\circ}$ cones from the vertex position, centered around each PMT that was hit within a 20 ns
time window (after time of flight subtraction). Pairs of such cones are then used to define ``event direction candidates'',
which are vectors taken from the vertex position to the intersection points of the two projected cones on
the detector surface. Only cone pairs which intersect twice are taken as ``event direction candidates''. 
Clusters of these candidates are then found by forming vector sums which are within 50$^{\circ}$ of a ``central event direction''.
Once an ``event direction candidate'' has been used in the formation of a cluster, it is then not used as
a ``central event direction'' and is skipped in further vector sums. 
Further iterations of this process will use the vector sums as the ''central event directions'', serving to
maximize and center the clusters. After a couple of iterations, the vector sum with the largest magnitude is kept
as the ``best fit direction''. The multiple scattering goodness (MSG) is then defined by taking the magnitude of
the largest vector sum (the ``best fit direction'') and normalizing it by the number of ``event directions'' which
would result using all hit PMTs within the 20 ns time window. For example, a MSG value of 0.4 would mean
that 40$\%$ of all ``event directions'' based on hit PMTs within 20 ns are included in the vector sum. 

\section{Analysis Results}
The start of physics data taking occurred on October 6th,
2008, with this paper including data taken until December 31st, 2012.
The total livetime is 1306.3 days. The entire data period was taken using the same low energy threshold, 
with 84$\%$ triggering efficiency at 3.5-4.0 MeV, 
99$\%$ at 4.0-4.5 MeV and 100$\%$ above 4.5 MeV kinetic energy.

\begin{figure}[t]
  \centering
  \includegraphics[width=0.4\textwidth]{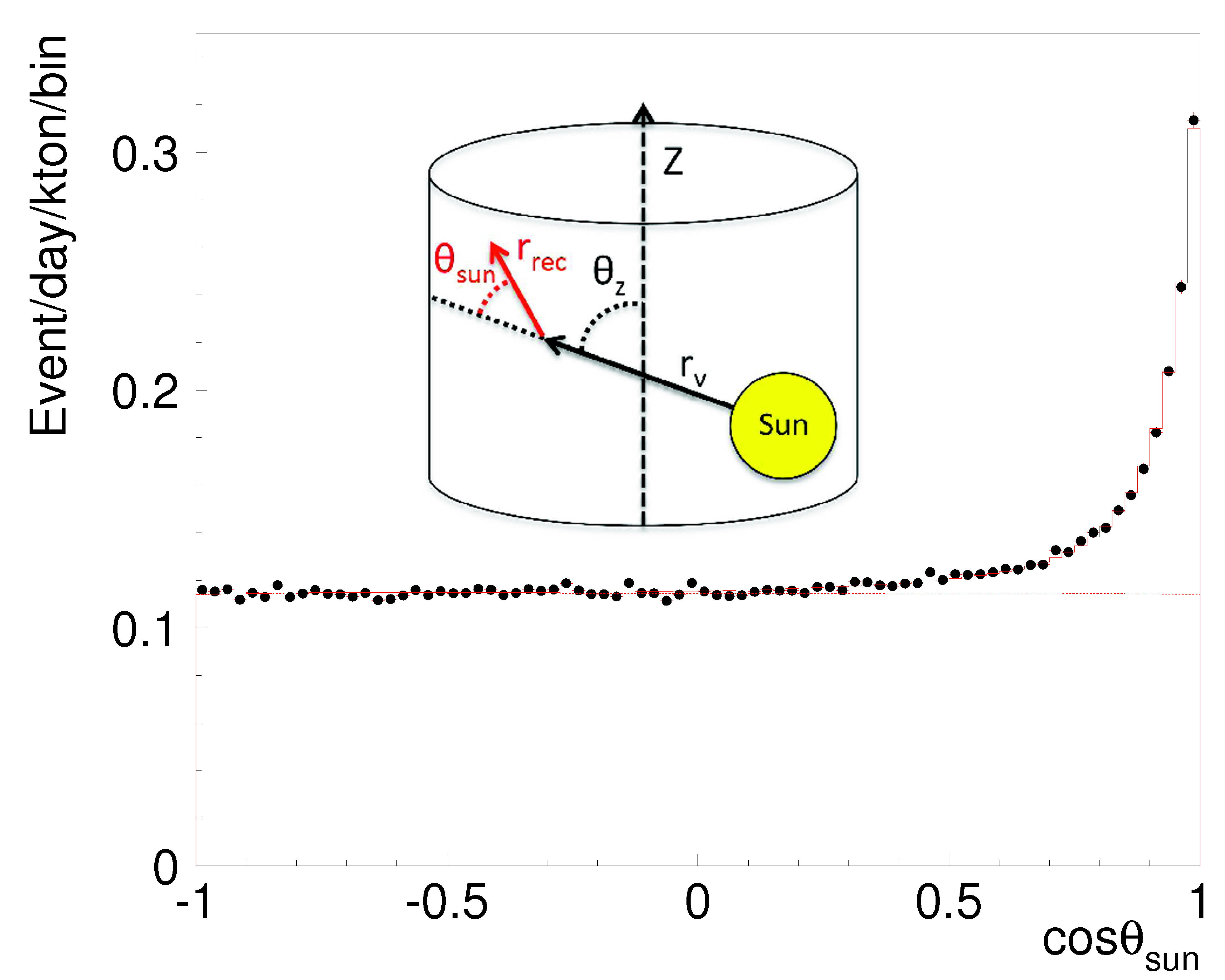}
  \caption{Solar angle distribution for 3.5-19.5 MeV. $\theta_{\rm sun}$ is
the angle between the incoming neutrino direction $r_{\nu}$ and the  reconstructed recoil electron direction $r_{\rm rec}$. 
($\theta_z$ is the solar zenith angle). Black points are data while the solid and dashed histograms are best fits to the 
background and signal plus background, respectively.}
  \label{fig:solarangle}
\end{figure}
In the case of $\nu$-e interactions of solar neutrinos in SK,
the incident neutrino and recoil electron directions are
highly correlated.  Fig.\ref{fig:solarangle} shows the $\cos \theta_{\rm sun}$ distribution
for events between 3.5-19.5 MeV as well as the definition
of $\cos \theta_{\rm sun}$. In order to obtain the number of solar
neutrino interactions, an extended maximum likelihood
fit is used. This method is also used in the SK-I~\cite{bib:1}, II~\cite{bib:5},
and III~\cite{bib:6} analyses. The solid line of Fig.\ref{fig:solarangle} is the best fit
to the data. The dashed line shows the background component
of that best fit. 
SK-IV has $N_{\rm bin}=22$ energy bins; 19 bins of 0.5 MeV width between
3.5-13.5 MeV, two energy bins of 1 MeV between 13.5
and 15.5 MeV, and one bin between 15.5 and 19.5 MeV. 
Below 7.5 MeV, each bin is split into three sub-samples of MSG, with boundaries set at
MSG=0.35 and 0.45. These three sub-samples are then
fit simultaneously to a single signal and three independent background components.

\begin{figure}[th]
  \centering
  \includegraphics[width=0.5\textwidth, angle=0]{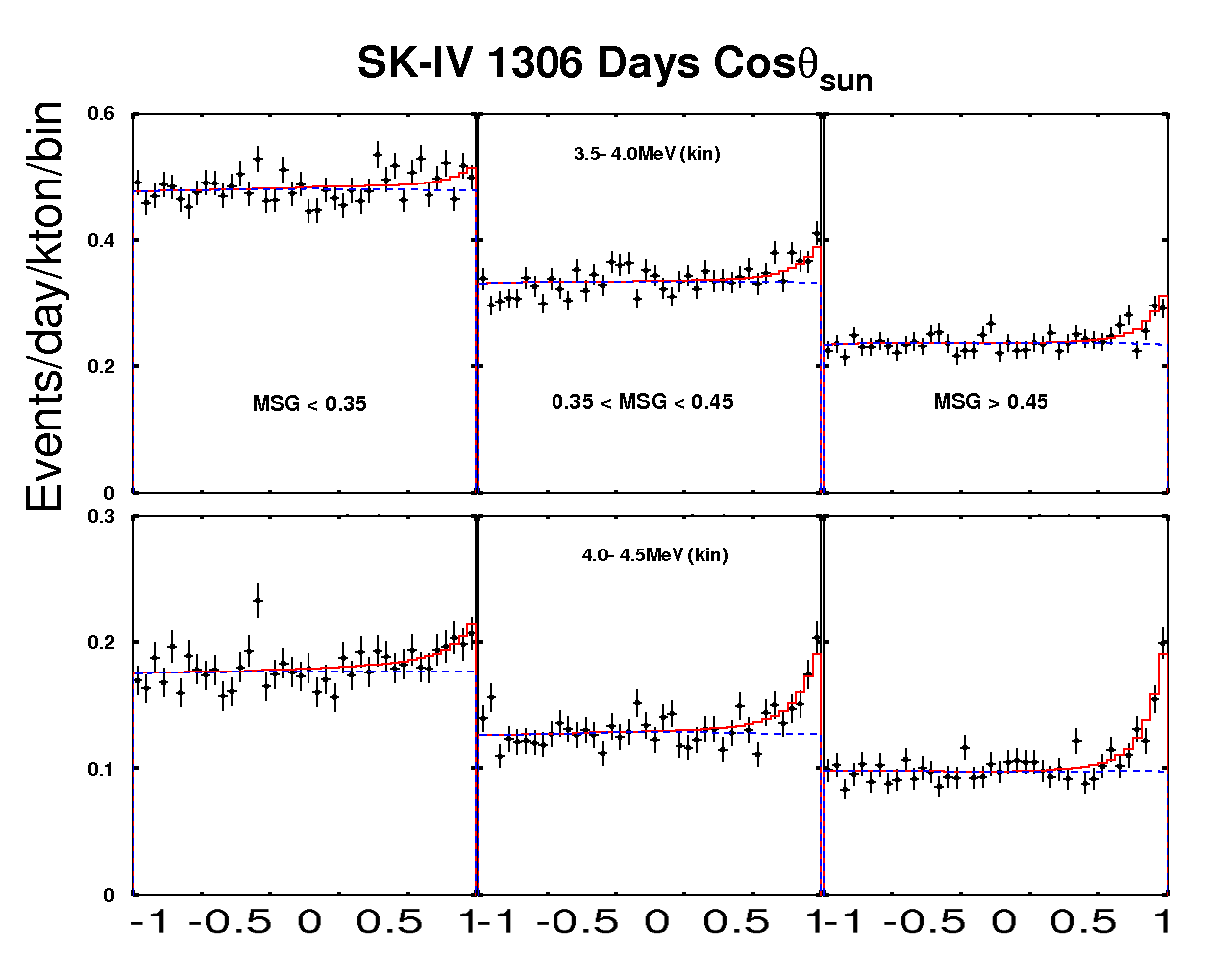}
  \caption{$\cos \theta_{\rm sun}$ for the two lowest (3.5-4.0 and 4.0-4.5 MeV) energy bins (upper and lower), 
for each MSG bin (left to right). Black points show the data while the blue and red histograms show the best fit to the
background and signal plus background, respectively.}
  \label{fig:solarMSG}
\end{figure}

Fig.\ref{fig:solarMSG} shows the measured angular distributions (as well as the fits) of
the lowest two (3.5-4.0 and 4.0-4.5 MeV) kinetic recoil electron energy bins 
for each MSG bin. As expected in the lowest energy bins, where the dominant
part of the background is due to very low energy gamma, beta decays, the background component is largest in the lowest
MSG sub-sample. Also, the solar neutrino elastic scattering peak sharpens as MSG is increased (and multiple
Coulomb scattering decreases). Using this method for recoil electron energy bins below 7.5 MeV gives 10$\%$
improvement on the statistical uncertainty of the number of signal events.

\begin{figure}[bh]
  \centering
  \includegraphics[width=0.45\textwidth]{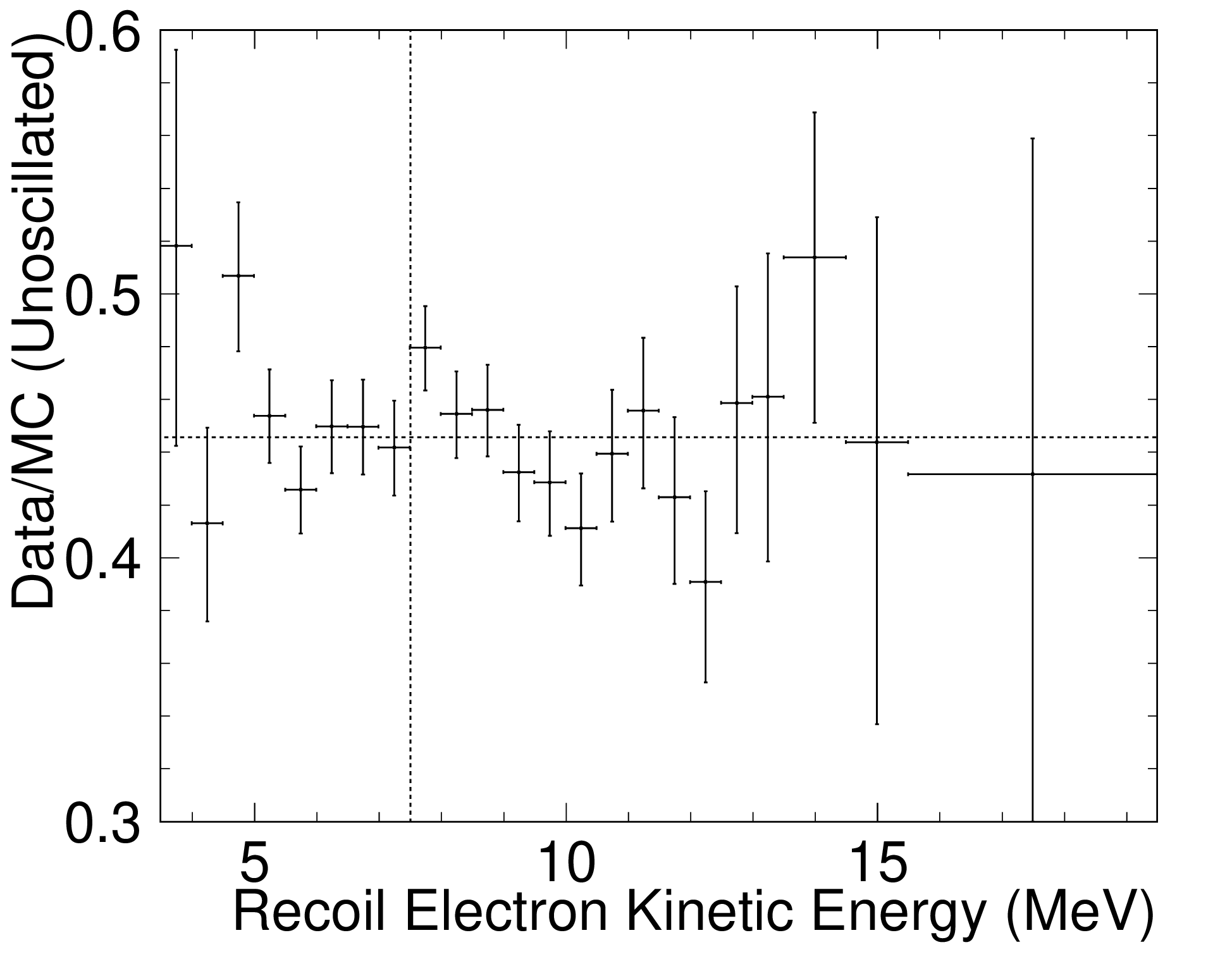}
  \caption{SK-IV energy spectrum using MSG below 7.5 MeV.
The horizontal dashed line gives the SK-IV total flux average
($2.36\times 10^6/({\rm cm}^2{\rm sec})$). Error bars shown are statistical plus energy-uncorrelated systematic uncertainties.}
  \label{fig:solarSpec}
\end{figure}

The combined systematic uncertainty of the total flux in SK-IV is found to be 1.7$\%$ as the quadratic sum of all components.
This is the best value seen throughout all phases of SK,
much improved over 2.2$\%$ in SK-III. The main contributions to the reduction come from improvements in the
uncertainties arising from the energy-correlated uncertainties (energy scale and resolution), the vertex shift,
trigger efficiency and the angular resolution. SK-III data below 6.0 MeV recoil electron kinetic energy has only
about half the livetime as the data above, while SK-IV's livetime is the same for all energy bins. As a consequence,
the energy scale and resolution uncertainties lead to a smaller systematic uncertainty of the flux in SK-IV
than in SK-III. The addition of the 3.5 to 4.5 MeV data lessens the impact of energy scale and resolution uncertainty 
on the flux determination even further. The number of solar neutrino events (between 3.5 and 19.5 MeV) is
$25,222^{+252}_{-250}(\rm{stat.})\pm429(\rm{syst.})$. This number corresponds
to a $^8$B solar neutrino flux of
$\Phi_{^8B} = (2.36 \pm 0.02({\rm stat.}) \pm 0.04({\rm syst.})) \times 10^6/({\rm cm}^2{\rm sec})$,
assuming a pure $\nu_e$ flavor content.
Fig.\ref{fig:solarSpec} shows the resulting SK-IV energy spectrum, where below
7.5 MeV MSG has been used and above 7.5 MeV the standard signal extraction method without MSG is used.

To test the expected ``upturn'' distortion below $\sim$6 MeV from the MSW resonance effects, 
energy-dependent parameterized functions of the $\nu_{e}$ survival probability ($P_{ee}$) were fitted to
all the SK-I to SK-IV spectra like SNO performed~\cite{SNO,SKIVsol}.
The fitting result shows that SK spectra disfavors flat suppression based on the constant $P_{ee}$ by $\sim1\sigma$ and
favors the ``upturn'' with also $\sim1\sigma$ significance.

The SK-IV livetime during the day (night) is 626.4 days (679.9 days). The solar neutrino flux between 4.5 and 19.5 MeV
and assuming no oscillations is measured as
$\Phi_D=(2.29\pm 0.03({\rm stat.})\pm 0.05({\rm sys.}))\times 10^6 /({\rm cm}^2{\rm sec})$ during
the day and $\Phi_N=(2.42 \pm 0.03({\rm stat.})\pm 0.05({\rm sys.}))\times 10^6 /({\rm cm}^2{\rm sec})$ during the night. 

A more sophisticated method to test the day/night effect is given in~\cite{bib:1,bib:20}. For a given set of oscillation parameters,
the interaction rate as a function of the solar zenith angle is predicted. Only the shape of the calculated solar zenith angle variation is used, the amplitude of it is scaled by an arbitrary parameter. The extended maximum likelihood fit to extract the solar neutrino signal is expanded to allow time-varying signals. The likelihood is then evaluated as a function of the average signal rates, the background rates and the
scaling parameter which is called the ``day/night amplitude''.
The equivalent day/night asymmetry is calculated by multiplying the fit scaling parameter with the expected day/night asymmetry. In this manner the day/night asymmetry is measured more precisely statistically.
Because the amplitude fit depends on the assumed
shape of the day/night variation, it necessarily depends on the oscillation
parameters, although with very little dependence expected on the mixing angles (in or near the large mixing angle solutions and for $\theta_{13}$ values consistent with reactor
neutrino measurements~\cite{bib:21}). 

\begin{figure}[t]
  \centering
  \includegraphics[width=0.43\textwidth, angle=0]{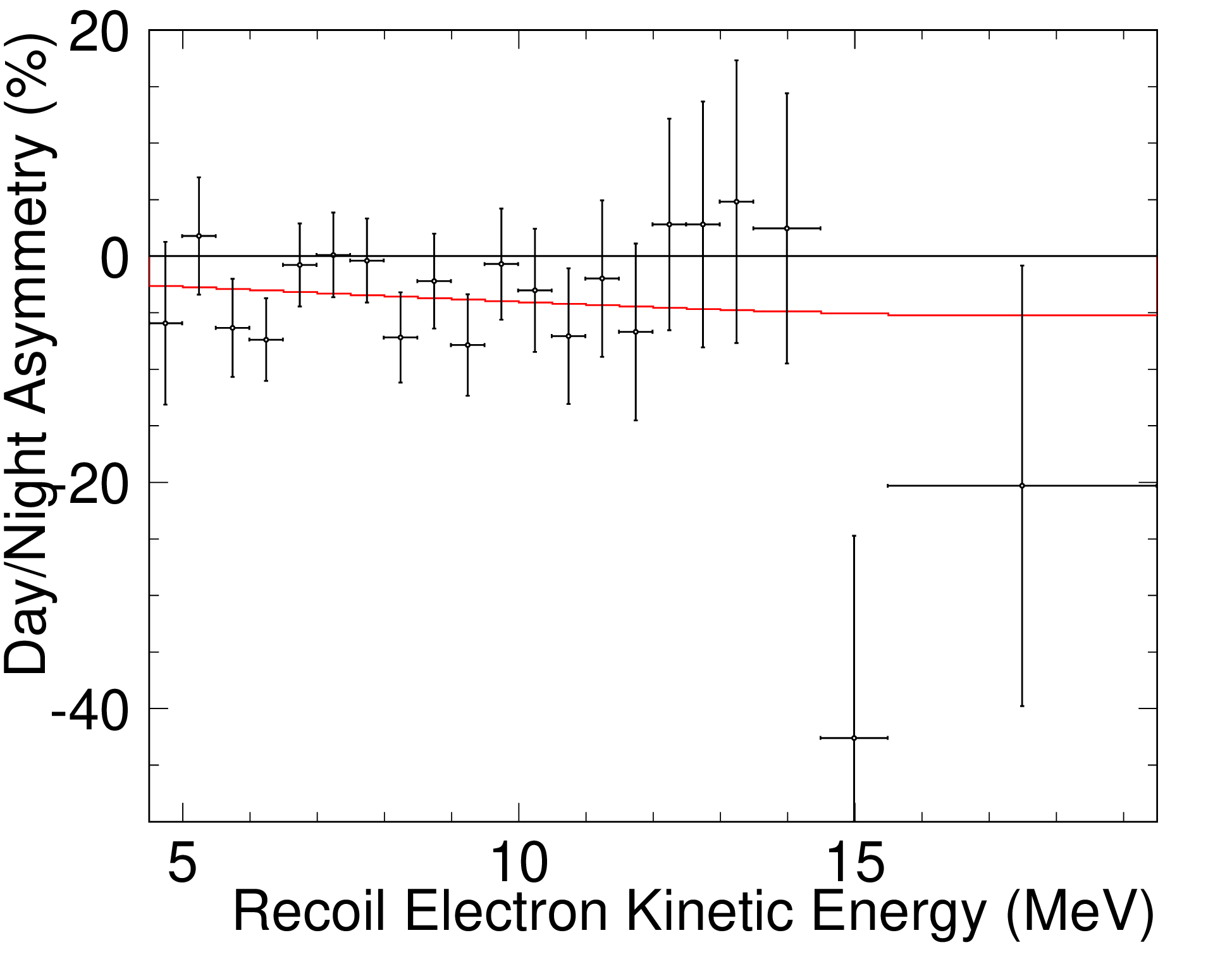}
  \caption{SK combined energy dependence of the fitted day/night asymmetry (measured day/night amplitude times the expected asymmetry (red)) for 
$\Delta m^2_{21}=4.89\times10^{-5} {\rm eV}^2$,
$\sin^2 \theta_{23}=0.314$ and $\sin^2 \theta_{13}=0.025$. The error bars shown are statistical uncertainties only.}
  \label{fig:solarDNspe}
\end{figure}
The day/night asymmetry coming from the SK-I to IV combined amplitude fit can be seen
as a function of recoil electron kinetic energy in Fig.\ref{fig:solarDNspe}, for $\Delta m^2_{21}=4.89\times 10^5 {\rm eV}^2$, $\sin^2_{12}=0.314$ and $\sin^2_{13}=0.025$. The day/night asymmetry in this figure is found by multiplying the fitted day/night amplitude
from each energy bin, to the expected day/night asymmetry (red distribution) from the corresponding bin.

\begin{figure}[bt]
  \centering
  \includegraphics[width=0.43\textwidth, angle=0]{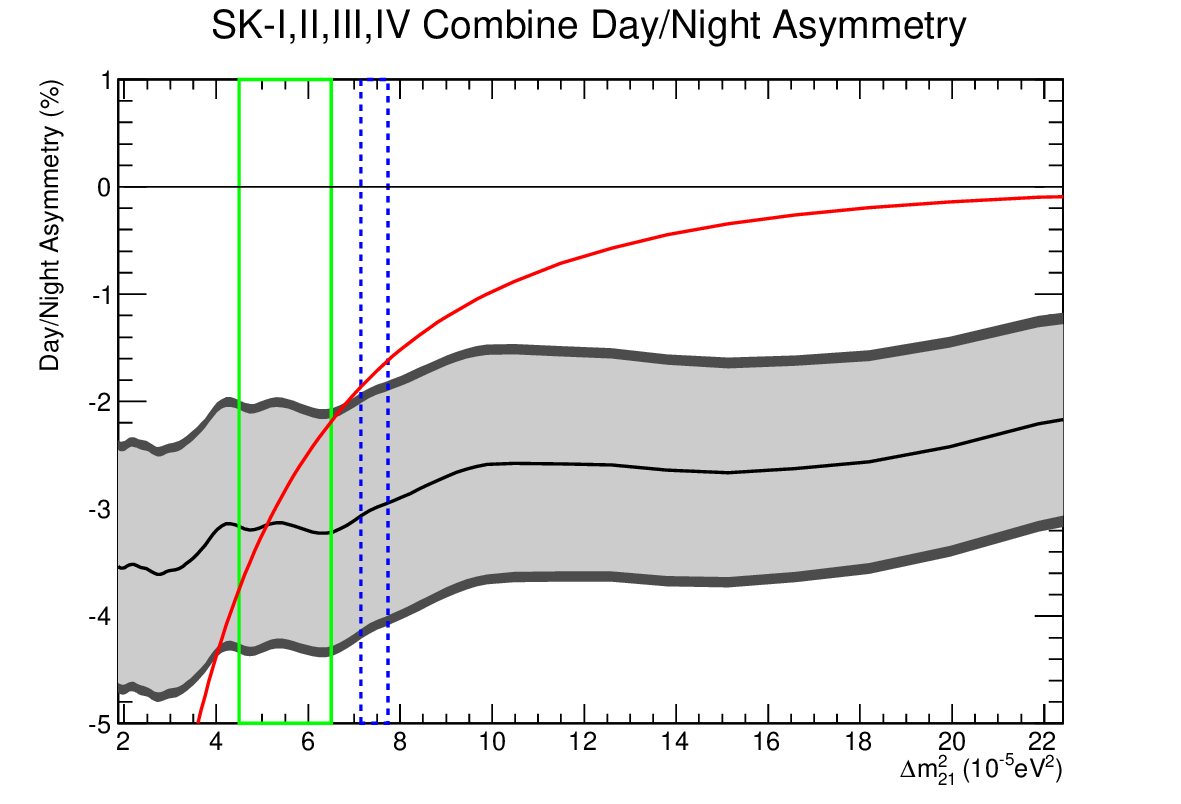}
  \caption{Dependence of the measured day/night asymmetry (fitted day/night amplitude times the expected day/night asymmetry (red)) on $\Delta m^2_{21}$ (light gray band=stat. error, dark gray band=stat.+syst. error) for $\sin^2 \theta_{12}=0.314$ and
$\sin^2 \theta_{13}=0.025$. Overlaid are the allowed ranges from solar neutrino data (green band) and KamLAND (blue band).}
  \label{fig:solarDNmass}
\end{figure}

Fig.\ref{fig:solarDNmass} shows the $\Delta m^2_{21}$ dependence of the SK all
phases combined day/night asymmetry for $\sin^2 \theta_{12}=0.314$ and $\sin^2\theta_{13}=0.025$. Here the day/night asymmetry is found by multiplying the fitted day/night amplitude by the expected day/night asymmetry (red curve). The point where the best fit crosses the expected curve represents the value of $\Delta m^2_{21}$ where the measured day/night asymmetry is equal to the expectation. Superimposed are the allowed ranges in $\Delta m^2_{21}$ from the global solar neutrino data fit (green) and from KamLAND (blue). The amplitude fit shows no dependence on the values of
$\theta_{12}$ (within the LMA region of the MSW plane) or $\theta_{13}$.

\section{Oscillation Analysis}
We analyzed the SK-IV elastic scattering rate, the recoil electron spectral shape and the day/night variation to constrain the solar neutrino oscillation parameters. The combination of SK-I, II, III and IV solar neutrino data measure the solar mixing angle
to $\sin^2\theta_{12}=0.341^{+0.029}_{-0.025}$ and the solar neutrino mass splitting 
to $\Delta m^2_{21} = 4.8^{+1.8}_{-0.9}\times 10^{-5}{\rm eV}^2$.

\begin{figure}[bth]
  \centering
  \includegraphics[width=0.42\textwidth, angle=0]{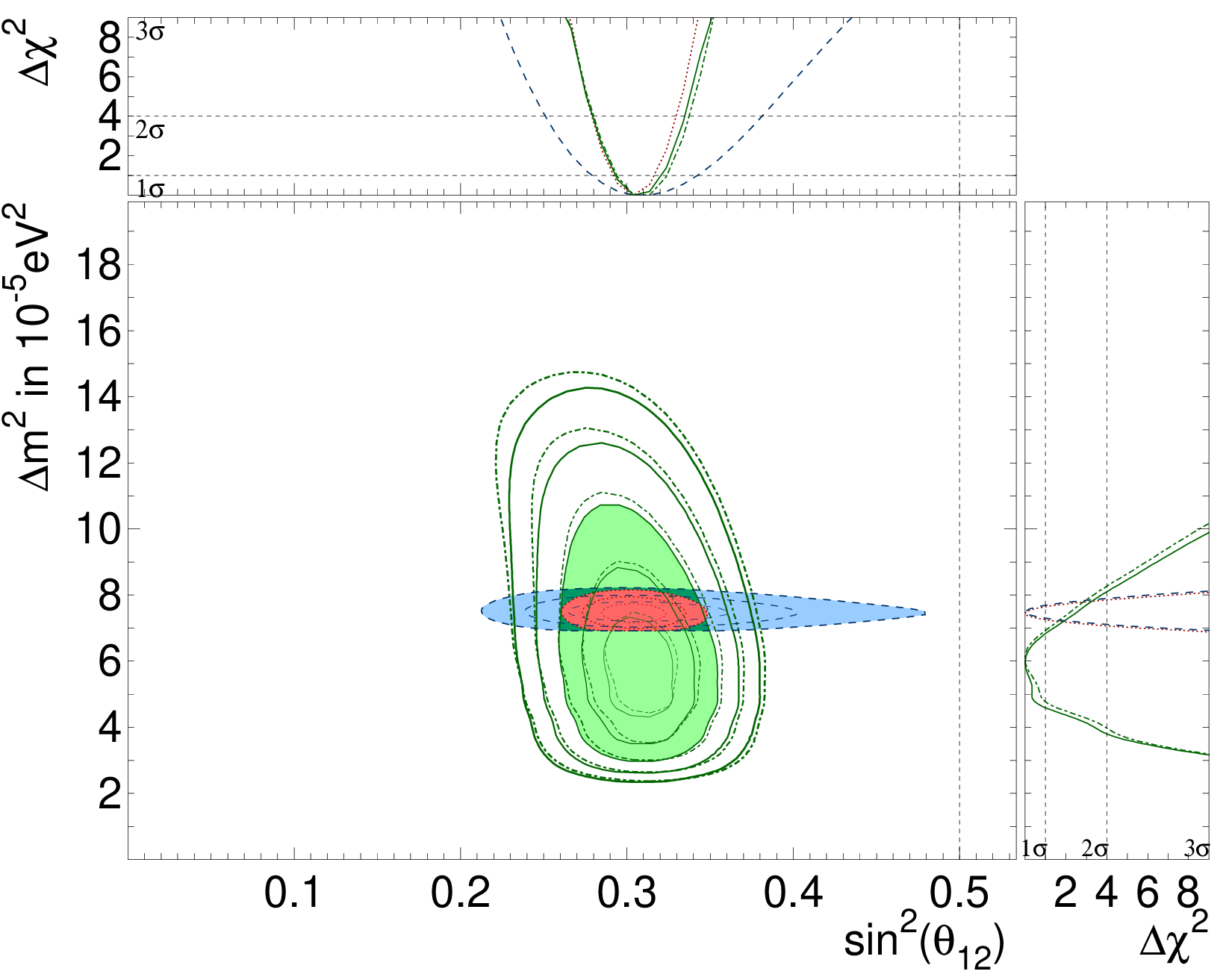}
  \caption{Allowed contours of $\Delta m^2_{21}$ vs. $\sin^2 \theta_{12}$ from solar
neutrino data (green) at 1, 2, 3, 4 and 5 $\sigma$ and KamLAND data (blue) at the 1, 2 and 3 $\sigma$ confidence levels.
Also shown are the combined results in red. For comparison,
the almost identical results of the SK+SNO combined fit are shown by the dashed dotted lines. $\theta_{13}$  is constrained by $\sin^2 \theta_{13} = 0.0242 \pm 0.0026$.}
  \label{fig:solar38}
\end{figure}

\begin{figure}[bth]
  \centering
  \includegraphics[width=0.42\textwidth, angle=0]{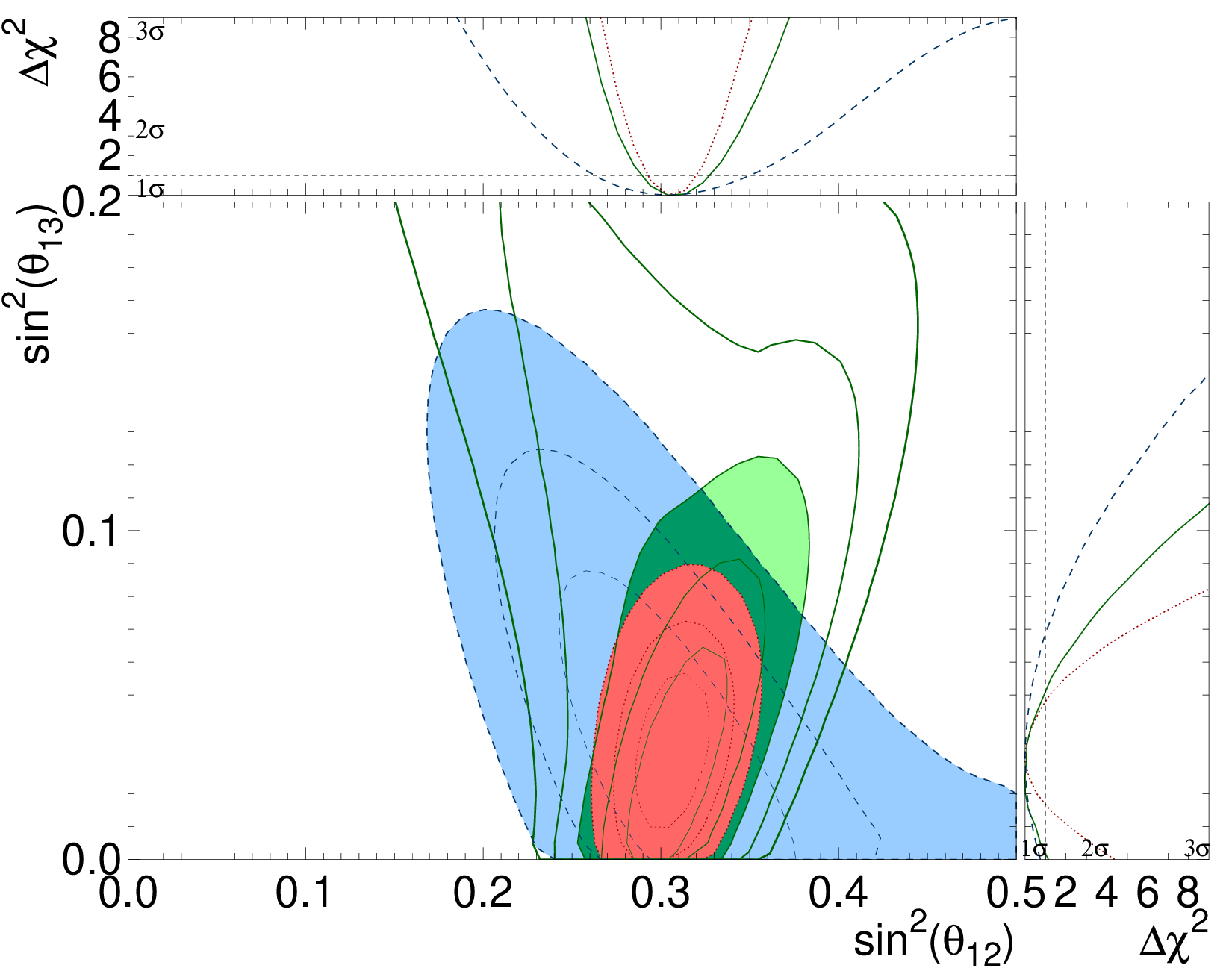}
  \caption{Allowed contours of $\sin^2 \theta_{13}$  vs. $\sin^2 \theta_{12}$ from solar neutrino data (green) at 1, 2, 3, 4 and 5 $\sigma$ and KamLAND measurements (blue) at the 1, 2 and 3 $\sigma$ confidence levels. Also shown are the combined results in red.}
  \label{fig:solar39}
\end{figure}

We then combined the SK-IV constraints with those of previous SK phases, as well as other experiments.
The allowed contours of all solar neutrino data (as well as KamLAND's constraints) are shown in Fig.\ref{fig:solar38} and \ref{fig:solar39}. In Fig.\ref{fig:solar38}  the contours from the fit to all solar neutrino
data are almost identical to the ones of the SK+SNO combined fit. In figures some tension between the solar neutrino and reactor anti-neutrino measurements
of the solar $\Delta m^2_{21}$ is evident. This tension is mostly due to the SK day/night measurement. Even though the expected amplitude agrees well within 1$\sigma$with the fitted amplitude for any $\Delta m^2_{21}$ in either the KamLAND or the SK range, the SK data somewhat favor the shape of the
variation predicted by values of $\Delta m^2_{21}$ that are smaller
than KamLAND's. In Fig.\ref{fig:solar39}, the significance of non-zero $\theta_{13}$ from
the solar+KamLAND data combined fit is about 2$\sigma$.

\section{Conclusion}
In the fourth phase of SK we measured the solar $^8$B neutrino-electron elastic scattering rate with the highest precision yet, (2.36$\pm$0.02(stat.)$\pm$0.04(syst.))$\times 10^6$ /(cm$^2$sec).
We find a 2.7 $\sigma$ indication for the existence of a solar day/night effect in the SK solar neutrino data, measured as the solar neutrino elastic scattering day/night rate asymmetry of $-3.2\pm1.1$(stat.)$\pm$0.5(syst.)$\%$.
SK's solar zenith angle variation
data results in the world's most precise measurement
of $\Delta m^2_{21} = 4.8^{+1.8}_{-0.9} \times 10^{-5} {\rm eV}^2$, using neutrinos rather than antineutrinos.
A fit to all solar neutrino data and KamLAND yields
$\sin^2 \theta_{12} = 0.304 \pm {0,013}$, $\sin^2 \theta_{13} = 0.031^{+0.017}_{-0.015}$ and 
$\Delta m^2_{21} = 7.45^{+0.20}_{-0.19} \times 10^{-5} {\rm eV}^2$. 
This value of $\theta_{13}$ is in agreement with reactor neutrino measurements.


\begin{thebibliography}{}

\bibitem{bib:1} J.Hosaka et al., Phys. Rev. D73, 112001 (2006).
\bibitem{bib:2} Q.R.Ahmad et al., Phys. Rev. Lett. 87 071301 (2001).
\bibitem{bib:4} S.P.Mikheyev and A.Y.Smirnov, Sov. Jour. Nucl. Phys. 42, 913 (1985); L.Wolfenstein, Phys. Rev. D17, 2369 (1978).
\bibitem{bib:5} J.P.Cravens et al., Phys. Rev. D 78, 032002(2008).
\bibitem{bib:6} K.Abe et al., Phys. Rev. D 83 052010 (2011).
\bibitem{bib:7} H.Nishino et al Nucl. Inst and Meth A.620(2009).
\bibitem{SNO}  B.Aharmim et al., Phys. Rev. C. 81, 055504 (2010).
\bibitem{SKIVsol} Super-Kamiokande Collaboration, Will be submitted to Phys. Rev. D.
\bibitem{bib:20} M.B.Smy et al., Phys. Rev. D. 69, 011104(R) (2004).
\bibitem{bib:21} F.P.An et al., arXiv:1210.6327 (2012); J.K.Ahn et al., Phys.Rev.Lett. 108 191802 (2012); Y.Abe et al.,Phys.Rev. D86 052008 (2012).

\end{thebibliography}
\end{document}